\pdfoutput=1
\documentclass{iau}
\usepackage{graphicx}
\usepackage{amsmath}
\usepackage{amssymb}
\usepackage[authoryear]{natbib}

\newcommand*{\be}{\begin{equation}}
\newcommand*{\ee}{\end{equation}}
\newcommand*{\bea}{\begin{eqnarray}}
\newcommand*{\eea}{\end{eqnarray}}
\newcommand*{\bme}{\begin{multiequations}}
\newcommand*{\eme}{\end{multiequations}}

\newcommand{\bm}[1]{\mbox{\boldmath$#1$\unboldmath}}

\newcommand{\aap}{Astron. Astrophys.}
\newcommand{\pre}{Phys. Rev. E}

\newcommand{\EMF}{\mbox{\boldmath ${\cal E}$} {}}

\title[]
{Helicity--vorticity turbulent pumping of magnetic fields in the solar
dynamo}

\author{V.~V.~Pipin}
\affiliation{
Institute Solar-Terrestrial Physics, Irkutsk, Russia
\\  email:pip@iszf.irk.ru}
\pubyear{2013}
\volume{294}  
\pagerange{--}
\jname{Solar and Astrophysical Dynamos and Magnetic
Activity}
\editors{A.G. Kosovichev, E.M. de Gouveia Dal Pino, \& Y.Yan, eds.}
\begin{document}

\maketitle
\begin{abstract}
The interaction of helical convective
motions and differential rotation in the solar convection zone results
in  turbulent drift of a large-scale magnetic field. We discuss the
pumping mechanism and its impact on the solar dynamo. 
\end{abstract}

\begin{figure}[ht]
\includegraphics[width=0.4\textwidth]{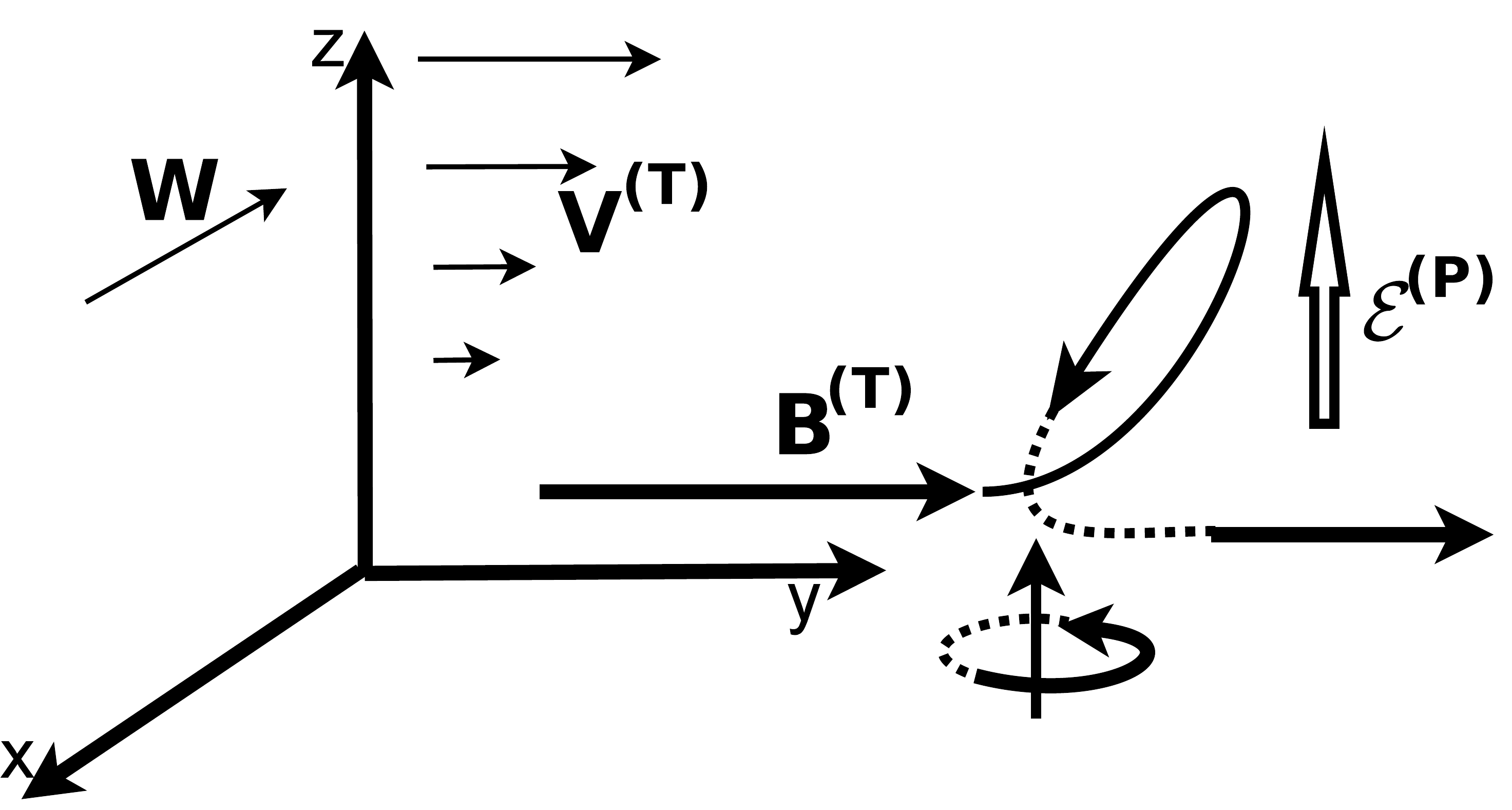}
\includegraphics[width=0.4\textwidth]{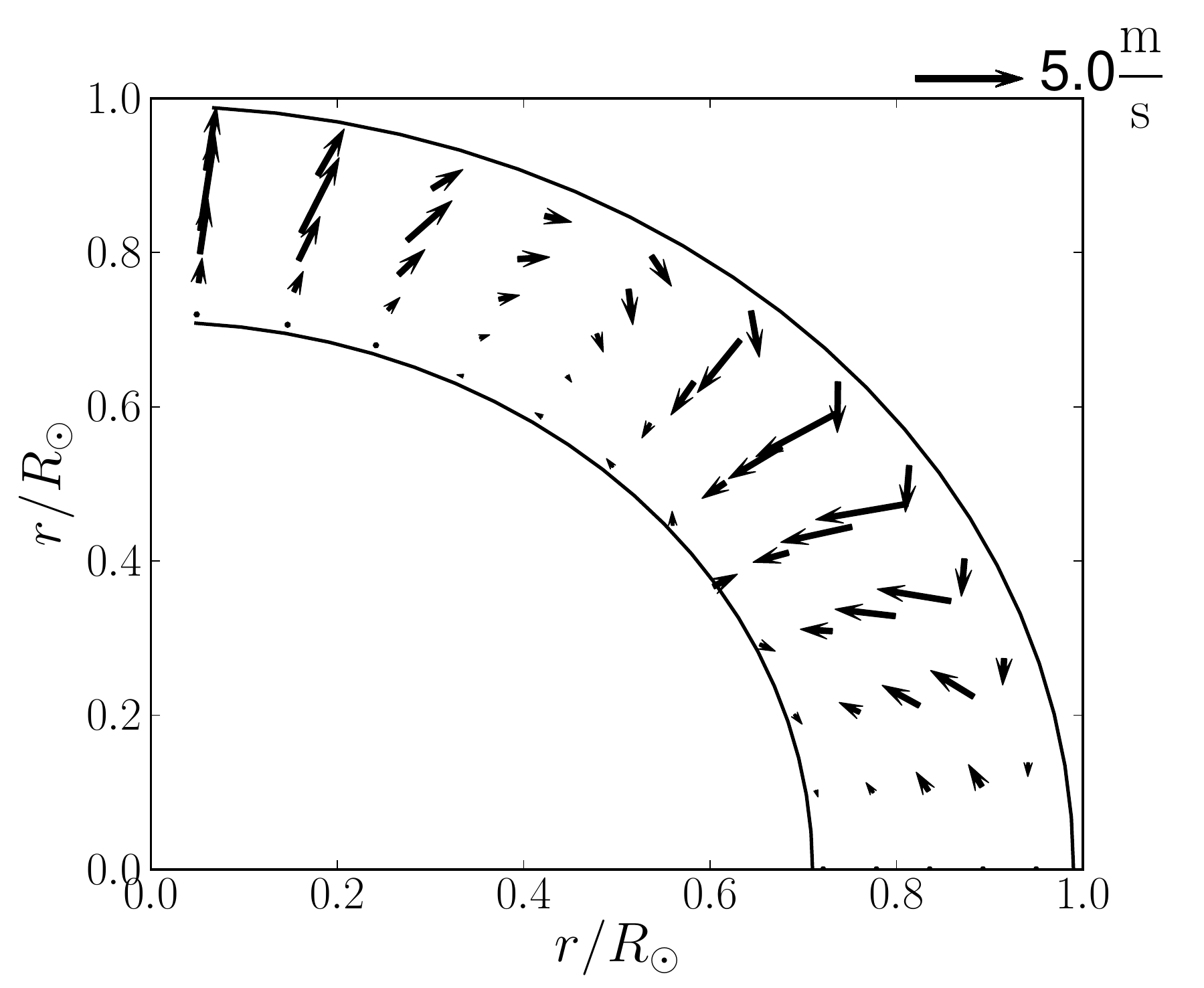}
\caption{\label{fig:The-field-line}The field lines of the large-scale magnetic
field, $\bm{B}^{(T)}$, are transformed by the helical motions
to a twisted $\varOmega$-like shape. This loop is folded by the large-scale
shear, $\bm{V}^{(T)}$, into the direction of the background large-scale
magnetic field, $\bm{B}^{(T)}$. The induced electromotive force
has a component, $\EMF^{(P)}$, which is perpendicular to
the field $\bm{B}^{(T)}$. The resulting effect is identical to
the effective drift of the large-scale magnetic field along the $x$-axis,
in the direction opposite to the large-scale vorticity vector $\bm{W}=\bm{\nabla}\times\bm{V}^{(T)}$,
i.e., $\EMF^{(P)}\sim-\bm{W}\times\bm{B}^{(T)}$.}
\end{figure}
\section{Introduction}

Recently \citep{pi08Gafd,mitr2009AA,2010GApFD.104..167L}, it has been
found that the helical convective motions and the helical turbulent
magnetic fields interacting with large-scale magnetic fields and
differential rotation can produce effective pumping in the direction
of the large-scale vorticity vector.
Figure \ref{fig:The-field-line}
illustrates the principal processes that induce the helicity--vorticity
pumping effect. 
A comprehensive study of the linear helicity--vorticity pumping effect for the case of weak
shear and slow rotation was given by \citet{garr2011} and their results
were extended by the direct numerical simulations (DNS) with a more general test-field method \citet{bran2012AA}.

\section{The solar dynamo model}

\begin{figure}
a)\includegraphics[width=0.85\textwidth,height=2.cm]{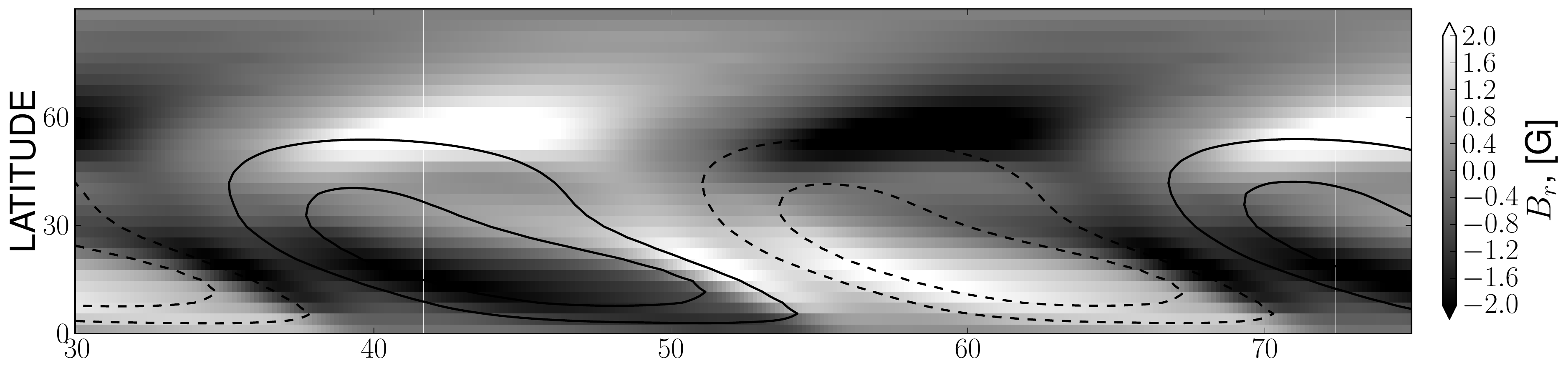}

b)\includegraphics[width=0.8\textwidth,height=2.cm]{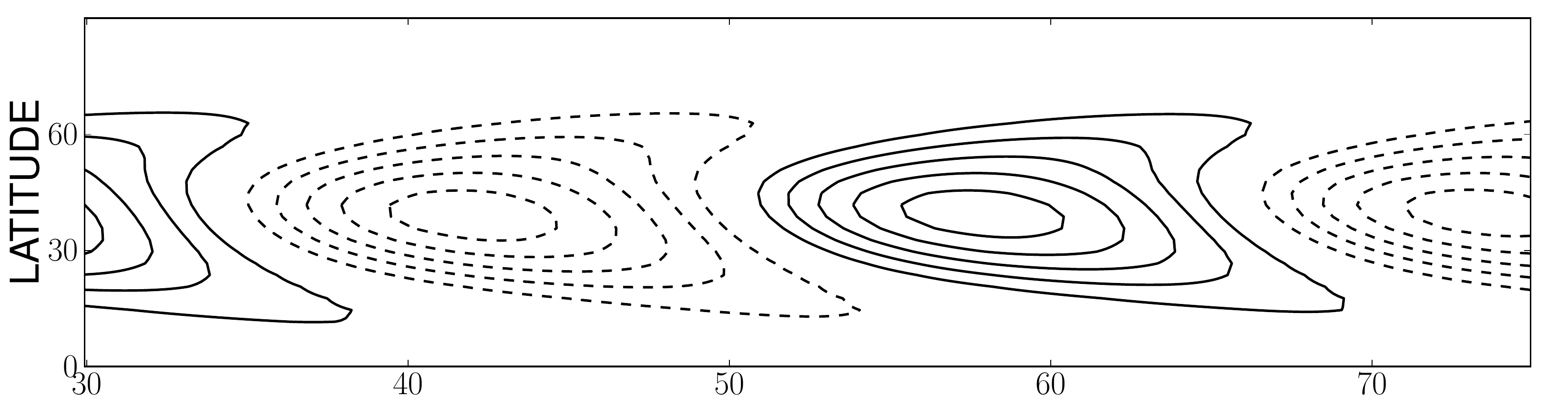}

c)\includegraphics[width=0.85\textwidth,height=2.cm]{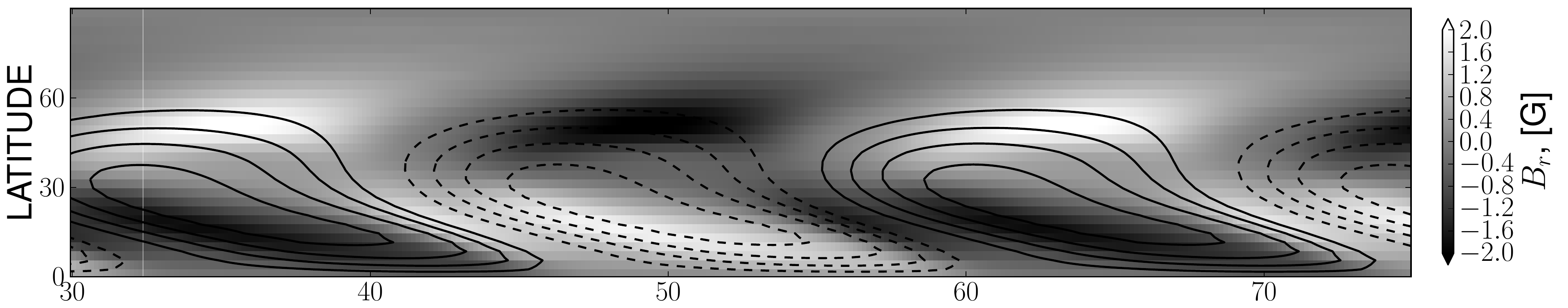}

d)\includegraphics[width=0.8\textwidth,height=2.cm]{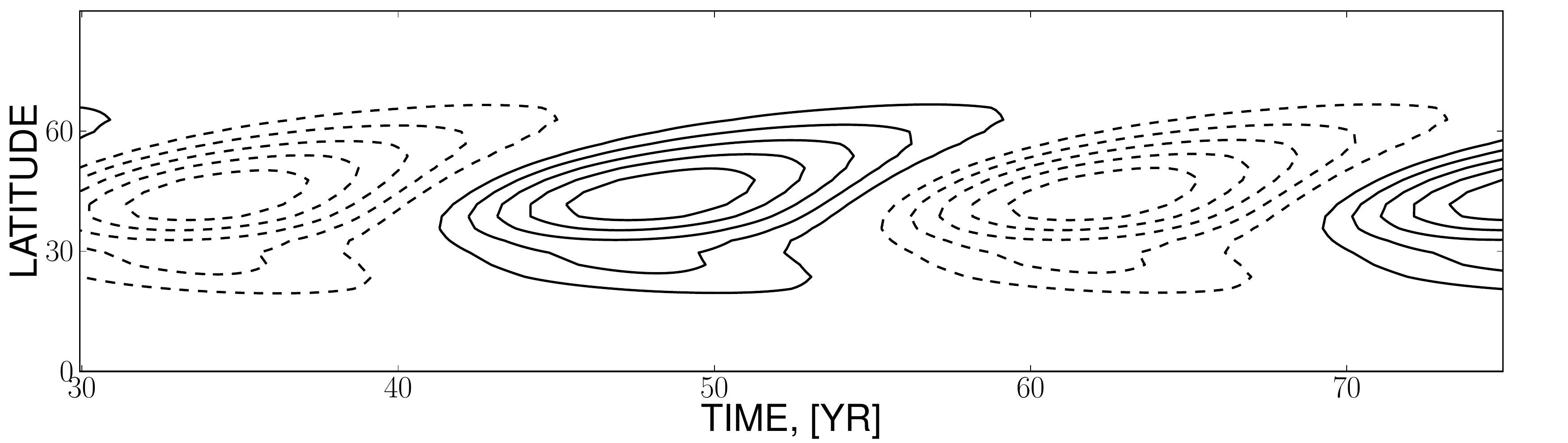}

\caption{\label{evo1}The time-latitude diagrams for the toroidal and radial
magnetic fields for the models D1 and D2: a) the model D1, the toroidal
field (iso-contours, $\pm.25KG$) near the surface and the radial field
(gray-scale density plot); b) the model D1, the toroidal field at
the bottom of the solar convection zone, the contours drawn in the
range $\pm.5KG$; c) the same as for item a) for the model D2; d)
the same as for item b) for the model D2.}
\end{figure}

It is found that the magnetic helicity contribution of the pumping
effect can be important for explaining the fine structure of the
sunspot butterfly diagram. In particular, the magnetic helicity contribution
results in a slow-down of equatorial propagation of the dynamo wave.
The slow-down starts just before the maximum of the cycle. 
For the time being it is unclear
what are the differences in predictions between different dynamo
models and how well do they reproduce the observations. A more detailed
analysis is needed.

{\bf Acknowledgments}
\medskip
I thank for the support the RFBR grants
12-02-00170-a, 10-02-00148-a and 10-02-00960,  the support of the  Integration Project of SB RAS
N 34, and  support of the state contracts 02.740.11.0576, 16.518.11.7065 of the Ministry
 of Education and Science of Russian Federation.

\end{document}